\documentclass[10pt]{iopart}

\usepackage{graphicx}

\usepackage{iopams}
\usepackage{mathrsfs}

\usepackage{xcolor}
\usepackage{braket}
\usepackage{bm}

\begin{document}
\title[Impact of spin--orbit coupling on orbital diamagnetism in a narrow-gap semiconductor $\mathrm{Pb}_{1-x}\mathrm{Sn}_x\mathrm{Te}$]{Impact of spin--orbit coupling on orbital diamagnetism in a narrow-gap semiconductor $\mathrm{Pb}_{1-x}\mathrm{Sn}_x\mathrm{Te}$}

\author{Yuki Mitani$^1$ and Yuki Fuseya$^{1,2}$}
\address{$^1$ Department of Engineering Science, the University of Electro-Communications, Chofu, Tokyo 182-8585, Japan}
\address{$^2$ Department of Physics, Kobe University, Kobe 657-8501, Japan}
\ead{ma004465@edu.cc.uec.ac.jp}

\begin{abstract}
We study the influence of spin--orbit coupling (SOC) on orbital magnetism in $\mathrm{Pb}_{1-x}\mathrm{Sn}_x\mathrm{Te}$, a narrow-gap semiconductor.
Using the $\pi$-matrix method, we calculate material-specific Landau levels and evaluate the magnetization, fully including interband effects.
The system exhibits diamagnetism for both $x = 0$ and $x = 0.35$, with the latter showing a stronger response due to its smaller gap.
The magnitude of diamagnetism increases monotonically with SOC strength, particularly in strong magnetic fields.
To clarify the underlying mechanism, we introduce the free--Zeeman--Dirac (fZD) model and fit its parameters to the calculated Landau levels.
The analysis reveals that SOC enhances the Dirac-type interband contribution relative to the Zeeman term, leading to increased diamagnetism.
These results demonstrate that SOC can play a key role in orbital magnetism through interband effects.
\end{abstract}


\submitto{\JPCM}

\ioptwocol

\section{Introduction}
Orbital diamagnetism is a long-studied phenomenon in solid-state physics, dating back to its early days~\cite{Peierls1933_LandauPeierlsFormula,FuseyaReviewDirac2015}.
Beyond its historical significance, orbital diamagnetism is closely linked to contemporary topics in condensed matter physics, such as the spin Hall effect ~\cite{FuseyaOgataFukuyama2012_SHEDiamagnetism_IsotropicWolff,Stredad2006_SHE_ChemicalPotentialGradient} and fundamental topics, \textit{e.g.}, the Hall effect~\cite{Ogata2024_OrdinalHallOrbitalSusceptibility}.
While the diamagnetism of conventional metals can be well understood by the Landau--Peierls theory~\cite{Peierls1933_LandauPeierlsFormula}, many materials—such as graphite~\cite{AkibaTokunaga2015_Graphite}, graphene~\cite{Fukuyama2007_2DWeylDiamagnetism,KoshinoAndo2010_DiamagnetismGrapheneWolff}, bismuth~\cite{OtakeMomiuchiMatsuno1980_SusceptibilityBismuth,Iwasa2019}, PbTe~\cite{AkimenkoStarykTovstyuk1967_MagnePbTe,AkibaTokunaga_PbTe2018} and other Dirac electron systems~\cite{PratamaUkhtarySaito2021_MagnetizationMassiveDirac,KeserLyanda-GellerSushkov2022_QEDDirac}—exhibit anomalously large diamagnetic responses that cannot be captured by Landau--Peierls theory.
A key to understanding these cases lies in the inclusion of interband magnetic effects, which are beyond the scope of the Landau--Peierls framework~\cite{FukuyamaKubo1969_SusceptibilityTwoBandModel}.

The importance of interband magnetic effects in diamagnetism has since been recognized for explaining giant diamagnetism~\cite{FuseyaReviewDirac2015,KuboFukuyama1970_ConfSemiconductors,BuotMcClure1972_BismuthDiamagnetism}.
The interband effects in diamagnetism have also been attracting attention for their close connection with broader concepts in condensed matter physics, including Berry curvature, topological properties~\cite{Xiao2010_RMP_BerryPhase,Ando2013_ReviewTopologicalInsulators,MikitikSharlai2007_BerryPhaseExpSummary,MikitikSharlai2019_SusceptTopolSemimetals}, and the spin-Hall effect~\cite{FuseyaReviewDirac2015,FuseyaOgataFukuyama2012_SHEDiamagnetism_IsotropicWolff,FuseyaOgataFukuyama2014_SHEDiamagnetism_AnisotropicDirac}.
Among the various factors influencing interband effects, spin--orbit coupling (SOC) is often considered essential---particularly in solids where relativistic effects play a non-negligible role.

However, the actual role of SOC in orbital diamagnetism has remained unclear.
On one hand, materials with strong SOC, such as bismuth and PbTe, tend to show significant diamagnetic responses, suggesting that SOC may enhance interband effects.
On the other hand, systems with negligible SOC, such as graphite and graphene, also exhibit enhanced diamagnetism.
These facts raise a fundamental and unresolved question: ``does SOC enhance or suppress orbital diamagnetism?"

A possible solution is recognizing that SOC can influence multiple competing contributions to magnetism.
As we will discuss later in section~\ref{Sec_fZD}, SOC generally modifies both Zeeman spin splitting, which tends to enhance \emph{paramagnetism}, and Dirac-type interband mixing, which tends to enhance \emph{diamagnetism}.
Therefore, whether SOC leads to an overall increase or decrease in diamagnetism is not apparent a priori and must be determined case by case.

One of the reasons this issue has remained unresolved is the difficulty of directly probing the correlation between SOC and diamagnetism in crystalline systems.
A precise evaluation of diamagnetism, particularly its interband contributions, requires the calculation of the magnetic-field-quantized energy spectrum, namely, the Landau levels.
Until recently, such calculations were only feasible for simplified effective models, such as free electrons or Dirac electrons~\cite{Solyom,Wolff1964_WolffModel}.
It had long been impractical to compute Landau quantization while fully accounting for the material-specific band structure.
Moreover, while SOC is in principle encoded in the Dirac Hamiltonian, the process of reducing it to an effective model tends to obscure this information, thereby rendering the connection between SOC and diamagnetism ambiguous.
On the other hand, the relativistic linear combination of atomic orbitals (LCAO) calculations do retain a clear correspondence between SOC and the Hamiltonian.
However, to evaluate diamagnetism within this framework, one must diagonalize the Hamiltonian and apply the Landau--Peierls formula, which, as mentioned earlier, cannot account for interband magnetic effects.

Recently, a novel method called the $\pi$-matrix formalism has been developed by extending the matrix-mechanics approach~\cite{Izaki-PiMatrix}.
This method allows computation of Landau quantization while fully incorporating a material's intrinsic band structure.
When combined with relativistic LCAO calculations, this approach offers two key advantages: (i) the influence of SOC is explicitly represented in the Hamiltonian, and (ii) the resulting diamagnetic response rigorously includes interband magnetic effects.
This method opens the door to a direct investigation of the correlation between SOC and diamagnetism.
The present work is the first to undertake such a study.

\section{Theory}
\subsection{Magnetization}

The orbital magnetization $\boldsymbol{M}$ of a system in thermal equilibrium can be obtained from the thermodynamic potential $\Omega(\boldsymbol{B})$ as
\begin{equation}
\boldsymbol{M} = - \frac{\nabla_{B} \Omega(\boldsymbol{B})}{V},
\label{eq:magnetization}
\end{equation}
where $V$ is the volume of the system and $\boldsymbol{B}$ denotes the magnetic field.
The grand potential $\Omega(\boldsymbol{B})$ is given by
\begin{equation}
\Omega(\boldsymbol{B}) = - k_B T\sum_{\ell,\boldsymbol{k}} \ln \left\{1 + e^{-\left[E_\ell(\boldsymbol{k}, \boldsymbol{B}) - \mu\right]/k_B T}\right\},
\label{eq:thermodynamic_potential}
\end{equation}
where $E_\ell(\boldsymbol{k}, \boldsymbol{B})$ is the energy of the $\ell$-th band at wave vector $\boldsymbol{k}$ under magnetic field $\boldsymbol{B}$, $\mu$ is the chemical potential, and $T$ being the temperature.
The summation is performed over all bands $\ell$ and wave vectors $\boldsymbol{k}$ in the Brillouin zone.

\subsection{Landau Levels}

In this study, we calculate the Landau levels of materials under magnetic fields using the $\pi$-matrix method~\cite{Izaki-PiMatrix}, which enables the incorporation of material-specific band structures in a gauge-invariant and numerically rigorous manner.

While the $\pi$-matrix method can, in principle, be applied to any band calculation method, we employ a relativistic LCAO model in this work.
This is because the LCAO Hamiltonian explicitly incorporates SOC~\cite{Chadi1977_RelativisticLCAO,Lent_PbTeBands}, furthermore, it allows us to continuously tune the strength of SOC, thereby enabling a systematic investigation of its role in orbital diamagnetism.

We adopt PbTe as a platform for investigating the influence of SOC on diamagnetism for the following reasons.
First, PbTe is known to exhibit clear orbital diamagnetism in experiments~\cite{AkimenkoStarykTovstyuk1967_MagnePbTe,AkibaTokunaga_PbTe2018,NimtzSchlicht_1983_LeadSaltMonograph}.
Second, the SOC strength in PbTe is moderate—not too weak and not too strong—making it an ideal system for probing the correlation between SOC and diamagnetism~\cite{Izaki-PiMatrix}.
Third, the applicability and accuracy of the $\pi$-matrix method have already been demonstrated for PbTe, yielding not only qualitative but also quantitative agreement with experimental results~\cite{AkibaTokunaga_PbTe2018,Izaki-PiMatrix}.

The procedure for calculating Landau levels is as follows~\cite{Izaki-PiMatrix}.
We begin with the relativistic LCAO Hamiltonians for PbTe developed by Lent {\it et al.}~\cite{Lent_PbTeBands}.
This Hamiltonian explicitly includes spin--orbit couplings for the cations ($\lambda_{\mathrm{c}}$) and the anions ($\lambda_{\mathrm{a}}$).
The basis includes $s$-, $p$-, and $d$-orbitals for both cations (Pb) and anions (Te), along with spin degrees of freedom, resulting in a $36 \times 36$ matrix Hamiltonian.
Next, we transform the LCAO Hamiltonian into the Luttinger--Kohn representation used in $\boldsymbol{k} \cdot \boldsymbol{p}$ theory~\cite{LuttingerKohn-LKRepresentation,WillatzenLewYanVoonKdotP}, which retains the same matrix dimension.
To incorporate a magnetic field, we replace the canonical momentum $\bm{p}$ with the kinematical momentum $\bm{\pi} = -i \hbar \bm{\nabla} + e \bm{A}$, where $e > 0$ is the elementary charge and $\bm{A}$ is the vector potential.
Then we obtain the Hamiltonian in the Luttinger--Kohn representation under a magnetic field as follows:
\begin{eqnarray*}
    \fl\hat{H}_{\mathrm{LK}} = \\
    \small{
    \pmatrix{
        \varepsilon_0 + \displaystyle\frac{\pi^2}{2m_e} & 0 & \bm{\pi}\cdot\bm{v}_{01}^{\uparrow\uparrow} & \bm{\pi}\cdot\bm{v}_{01}^{\uparrow\downarrow} & \cdots \cr
        0 & \varepsilon_0 + \displaystyle\frac{\pi^2}{2m_e} & \bm{\pi}\cdot\bm{v}_{01}^{\downarrow\uparrow} & \bm{\pi}\cdot\bm{v}_{01}^{\downarrow\downarrow} & \cdots \cr
        \bm{\pi}\cdot\bm{v}_{10}^{\uparrow\uparrow} & \bm{\pi}\cdot\bm{v}_{10}^{\uparrow\downarrow} & \varepsilon_1 + \displaystyle\frac{\pi^2}{2m_e} & 0 & \cdots \cr
        \bm{\pi}\cdot\bm{v}_{10}^{\downarrow\uparrow} & \bm{\pi}\cdot\bm{v}_{10}^{\downarrow\downarrow} & 0 & \varepsilon_1 + \displaystyle\frac{\pi^2}{2m_e} & \cdots \cr
        \vdots & \vdots & \vdots & \vdots & \ddots
    }
    }
\end{eqnarray*}
where $\varepsilon_m$ is the energy of $m$-th band at the band extremum $\bm{k}_0$, $\bm{v}_{nm}^{\sigma\sigma'}$ denotes the velocity matrix element at $\bm{k}_0$, and $m_e$ is the bare electron mass.
We then express $\bm{\pi}$ as a matrix operator $\hat{\bm{\pi}}$ acting on the Landau level basis, such that the commutation relation $\hat{\bm{\pi}} \times \hat{\bm{\pi}} = -e \hbar \bm{B}$ is preserved.
An explicit matrix representation for $\bm{\pi}$ when $\bm{B}\parallel z$ is, for example,
\begin{equation*}
    \pi_x = \sqrt{\frac{e\hbar B}{2}}
    \pmatrix{
        0 & 1 & 0 & 0 & \cdots \cr
        1 & 0 & \sqrt{2} & 0 & \cdots \cr
        0 & \sqrt{2} & 0 & \sqrt{3} & \cdots \cr
        0 & 0 & \sqrt{3} & 0 & \cdots \cr
        \vdots & \vdots & \vdots & \vdots & \ddots
    }.
\end{equation*}
We note that in the $\pi$-matrix method, the effect of the vector potential $\bm{A}$ is included as the magnetic field $\bm{B}$ converted by the commutation relation of the kinematical momentum operator $\hat{\bm{\pi}}$ in its matrix representation.
This transformation eliminates the need for assuming a specific gauge, which guarantees the gauge-invariance of our calculations.
As a result, the final Hamiltonian becomes a $36N \times 36N$ matrix when the Landau level index is truncated at $N$.
We ensure that $N$ is sufficiently large to guarantee numerical convergence.

To systematically explore the role of SOC, we introduce a scaling factor $\kappa_{\mathrm{soc}}$ and define the modified SOC strengths as
\begin{equation*}
 \tilde{\lambda}_{\mathrm{c}, \mathrm{a}} = \kappa_{\mathrm{soc}} \lambda_{\mathrm{c}, \mathrm{a}}.
\end{equation*}
Here, $\kappa_{\mathrm{soc}} = 1$ corresponds to the original Lent model with full SOC, while $\kappa_{\mathrm{soc}} = 0$ entirely removes SOC.
It should be noted that varying $\kappa_{\mathrm{soc}}$ also alters the band gap of the system.
To isolate the effects of SOC from those of the band gap's shift, we introduce an additional tuning parameter $\nu\ (\nu \ge 0)$ that rescales the on-site energy $E_{\mathrm{pc}}$ of the cation $p$-orbitals.
By adjusting $\nu$ such that the band gap is fixed at the original value, we can examine how variations in $\kappa_{\mathrm{soc}}$ affect the band structure and Landau levels, independent of gap-related influences.
This approach allows us to focus on the intrinsic correlation between SOC strength and orbital diamagnetism.

\subsection{Free--Zeeman--Dirac Model}
\label{Sec_fZD}

To clarify the fundamental physics governing the relationship between spin--orbit coupling (SOC) and orbital magnetization, we introduce the free--Zeeman--Dirac (fZD) model.
This phenomenological model is designed to approximate the Landau levels of narrow-gap semiconductors with SOC and allows an intuitive interpretation of the roles played by different physical terms.

The energy dispersion of the fZD model is given by
\begin{eqnarray}
\fl E_{n,s,\pm}(k_{\parallel}, B) = \pm\Biggl[ \left(n+\frac{1}{2}\right)C_{\mathrm{f}}B + C_{\parallel\mathrm{f}}k_{\parallel}^2 \nonumber\\
 +\frac{s}{2} C_{\mathrm{Z}}B
 + \sqrt{\mathit{\Delta}^2+2\mathit{\Delta} C_{\mathrm{D}} jB
 + C_{\parallel\mathrm{D}} k_{\parallel}^2} \Biggr],
\label{eq:fZD_model}
\end{eqnarray}
where $s = \pm 1$, $n \ge 0$ is an integer Landau index, $j = n + 1/2 + s/2$, and $k_{\parallel}$ is the wave vector component parallel to the magnetic field.
The positive (negative) sign corresponds to the conduction (valence) bands.

The fZD model contains three main contributions to the magnetic response:
(i) the free-electron Landau quantization, represented by the $C_{\mathrm{f}}$ and $C_{\parallel\mathrm{f}}$ terms;
(ii) the Zeeman splitting, characterized by the $C_{\mathrm{Z}}$ term; and
(iii) the Dirac-type coupling, encoded in $C_{\mathrm{D}}$ and $C_{\parallel\mathrm{D}}$.
In addition to these terms, the model includes a band gap parameter $\mathit{\Delta}$, which sets the energy separation between the conduction and valence bands at zero magnetic field.
While $\mathit{\Delta}$ itself does not vary with $B$, it strongly influences the overall Landau level structure and the resulting magnetization, especially near the band edges.
This model provides a good approximation for the Landau levels near the band edges and is particularly useful for obtaining a physical understanding of magnetic response in narrow-gap semiconductors.

For the lowest Landau levels ($n = 0$, $s = -1$) at $k_{\parallel} = 0$, the energy simplifies to $E = \pm \left(- C_{\mathrm{Z}} B / 2 + \mathit{\Delta} \right)$.
However, in the actual Landau level structure of narrow-gap semiconductors, the lowest levels exhibit nonlinearity beyond linear-in-$B$ behavior~\cite{VecchiMSDresselhaus1974, ZhuFuseyEmptyingBismuthValleys2017}.
To account for this, we approximate their energy as
\begin{equation*}
E_{0,-1,\pm}(B) = \pm \left( \mathit{\Delta} + C_{\mathrm{lin}} B + C_{\mathrm{sq}} B^2 \right),
\end{equation*}
where the additional quadratic term captures the observed deviations.
These deviations may arise from the mixing between the lowest conduction and valence Landau levels due to interband coupling~\cite{VecchiMSDresselhaus1974, ZhuFuseyEmptyingBismuthValleys2017}.

\begin{figure}[tp]
    \centering
    \includegraphics[width=0.85\columnwidth]{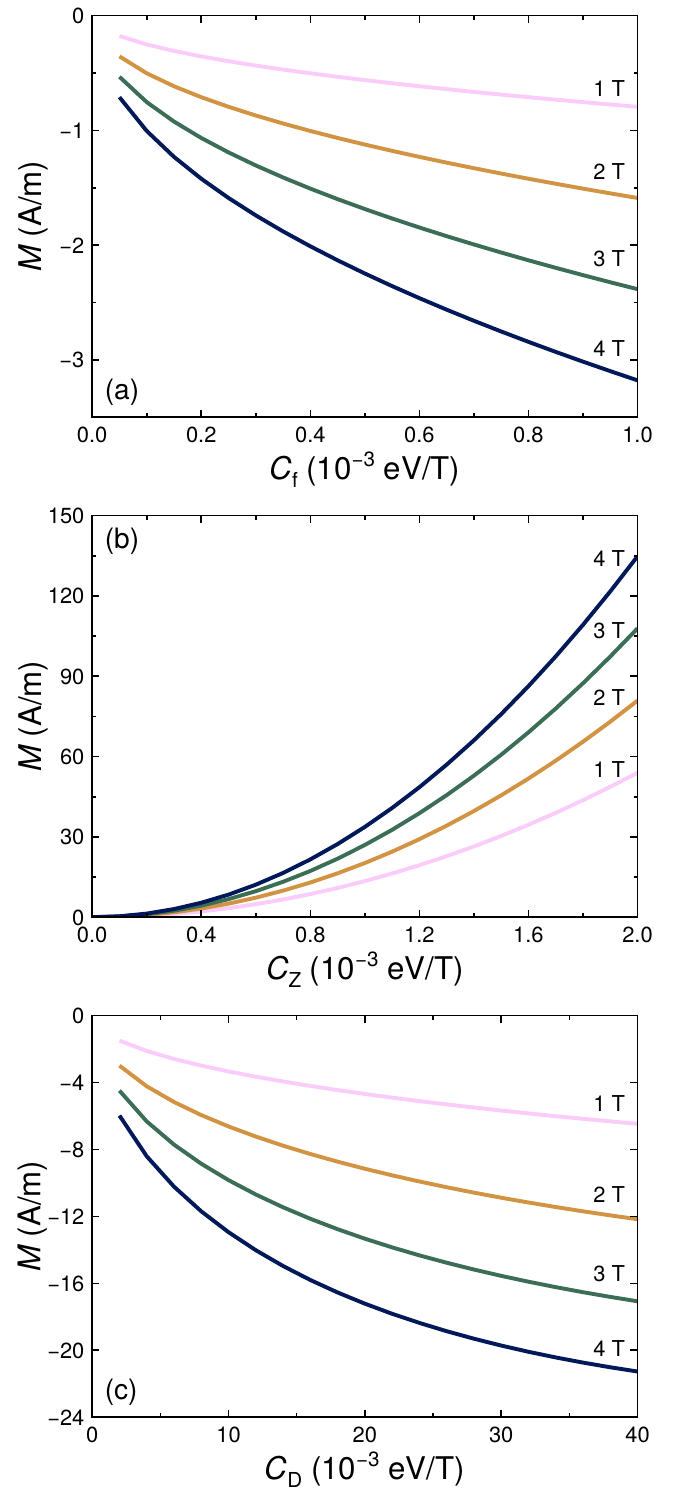}
    \caption{Contributions of the fZD model coefficients to the magnetization.
 (a) Magnetization with only the free-electron term $C_{\mathrm{f}}$ finite.
 (b) Additional contribution due to the Zeeman term $C_{\mathrm{Z}}$, with $C_{\mathrm{f}} = 5\times10^{-4}\ \mathrm{eV\,T^{-1}}$ fixed. The magnetization for $C_{\mathrm{Z}} = 0$ has been subtracted.
 (c) Magnetization with only the Dirac term $C_{\mathrm{D}}$ finite.
 A cutoff function is used in (c) to ensure convergence.}
    \label{fig:magnetization_fZD}
\end{figure}

When both valence and conduction bands are present in the energy spectrum, care must be taken to avoid unphysical contributions from high-energy states far away from the Fermi level, especially under strong magnetic fields.
To this end, we introduce an energy cutoff function $f_{\mathrm{c}}(\beta_{\mathrm{c}},E_{\mathrm{c}},E)$, defined as~\cite{KawamuraFuseya2023}
\begin{eqnarray}
f_{\mathrm{c}}(\beta_{\mathrm{c}},E_{\mathrm{c}}, E)
&= \frac{1}{4} \left[1+\tanh\left(\beta_{\mathrm{c}} (E_{\mathrm{c}} - E)\right)\right]
\nonumber\\&
\times \left[1+\tanh\left(\beta_{\mathrm{c}} (E + E_{\mathrm{c}})\right)\right],
\label{eq:cutoff_function}
\end{eqnarray}
where $E$ is the energy, $E_{\mathrm{c}}$ is the cutoff energy, and $\beta_{\mathrm{c}}$ controls the sharpness of the cutoff.
This function smoothly suppresses contributions from states with $|E| > E_{\mathrm{c}}$.
The introduction of this cutoff function is crucial for ensuring the convergence of $\Omega(\boldsymbol{B})$ and eliminating unphysical oscillations that may arise in strong magnetic fields, as discussed in Ref.~\cite{KawamuraFuseya2023}.

Figure~\ref{fig:magnetization_fZD} illustrates the contribution of each term in the fZD model to the magnetization. In these calculations, we set $\mathit{\Delta} = 0.05~\mathrm{eV}$ and $T = 77~\mathrm{K}$.
The lattice constant was taken as $a = 6.39 \times 10^{-10}~\mathrm{m}$. 
To ensure isotropic effective masses, we imposed the relation $C_{\parallel \mathrm{f}, \mathrm{D}} = (\hbar / 2e) C_{\mathrm{f}, \mathrm{D}} $.
Figure~\ref{fig:magnetization_fZD}(a) shows the magnetization when only $C_{\mathrm{f}}$ is finite. 
Figure~\ref{fig:magnetization_fZD}(b) shows the contribution from $C_{\mathrm{Z}}$.
(A finite value of $C_{\mathrm{f}}= 5\times10^{-4}~\mathrm{eV\,T^{-1}}$ was introduced to regularize the magnetization, which would otherwise diverge if only $C_{\mathrm{Z}}$ were present.)
 In both (a) and (b), only conduction bands are considered.
Figure~\ref{fig:magnetization_fZD}(c) displays the result when only $C_{\mathrm{D}}$ is non-zero, with both conduction and valence bands included.
A cutoff function [equation~(\ref{eq:cutoff_function})] was employed to ensure convergence.
The cutoff energy was set to $E_{\mathrm{c}} = 0.21~\mathrm{eV}$, and the smoothness parameter was $\beta_{\mathrm{c}} = 8~\mathrm{eV}^{-1}$.

These results reveal the distinct roles played by the free, Zeeman, and Dirac terms in the magnetization: the free term ($C_{\mathrm{f}}$) gives rise to Landau diamagnetism; the Zeeman term ($C_{\mathrm{Z}}$) contributes to Pauli paramagnetism, and the Dirac term ($C_{\mathrm{D}}$) leads to an additional diamagnetic contribution originating from the interband magnetic effect characteristic of Dirac electrons.

\begin{figure}[tbp]
    \centering
    \includegraphics[width=0.95\columnwidth]{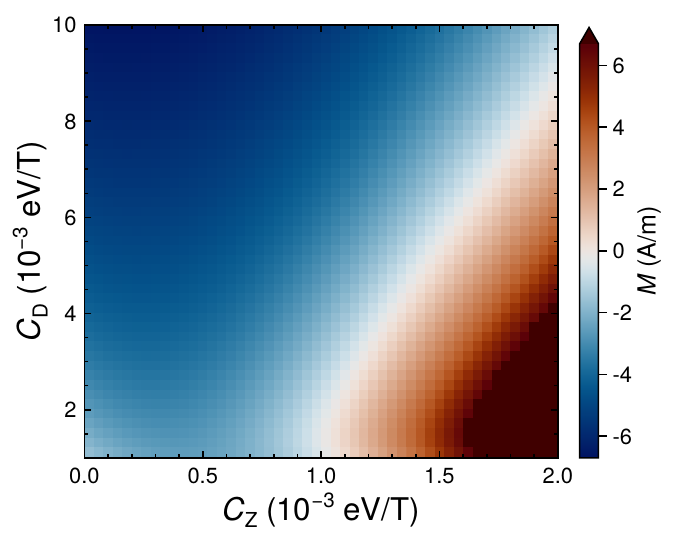}
    \caption{Magnetization as a function of the Zeeman ($C_{\mathrm Z}$) and Dirac ($C_{\mathrm D}$) coefficients within the fZD model at $B=2\,\mathrm{T}$.}
    \label{fig:magnetization_heatmap}
\end{figure}

Figure~\ref{fig:magnetization_heatmap} illustrates how the magnetization behaves when $C_{\mathrm{f}}$ is held fixed and $C_{\mathrm{Z}}$ and $C_{\mathrm{D}}$ are varied.
Here, we set $C_{\mathrm{f}} = 2 \times 10^{-4}~\mathrm{eV\,T^{-1}}$ and explored a wide range of values for $C_{\mathrm{Z}}$ and $C_{\mathrm{D}}$.
As the ratio $|C_{\mathrm{D}}/C_{\mathrm{Z}}|$ increases, the magnetization becomes more strongly diamagnetic.
This tendency confirms that the relative strength of the Dirac term to the Zeeman term is a key determinant of the magnetic behavior.
In these calculations, we used the following parameters:
band gap $\mathit{\Delta} = 0.05~\mathrm{eV}$, temperature $T = 77~\mathrm{K}$, cutoff energy $E_{\mathrm{c}} = 0.21~\mathrm{eV}$, smoothing parameter $\beta_{\mathrm{c}} = 8~\mathrm{eV}^{-1}$, and lattice constant $a = 6.39 \times 10^{-10}~\mathrm{m}$.
The effective mass was assumed to be isotropic.

In realistic multiband systems, SOC is generally expected to influence both $C_{\mathrm{Z}}$ and $C_{\mathrm{D}}$.
Specifically, SOC can (i) enhance the effective $g$-factor, leading to a larger $|C_{\mathrm{Z}}|$ and stronger paramagnetic response, and (ii) increase the Diracness of the band structure, thereby amplifying $C_{\mathrm{D}}$ and enhancing orbital diamagnetism.
The former is exemplified by bismuth, which exhibits unusually large and anisotropic $g$-factors~\cite{BiAngleResolvedLLPRB2011, Fuseya_HoleGFactor2015}.
The latter reflects the essential role of SOC in realizing Dirac electrons in solids, as discussed in Wolff’s theory~\cite{FuseyaReviewDirac2015,Wolff1964_WolffModel}.
Therefore, SOC may contribute to either paramagnetic or diamagnetic behavior, depending on which effect dominates.
This ambiguity underscores the importance of quantitatively analyzing how SOC affects each fZD coefficient in real materials—a central aim of the present study.


\section{Results and Discussion}
\subsection{Landau Levels and Magnetization in $\mathrm{Pb}_{1-x}\mathrm{Sn}_x\mathrm{Te}$}

In this subsection, we present the results of applying the method described in the previous section to the real material PbTe.
It is well known that the band gap of PbTe can be tuned by substituting Sn, forming $\mathrm{Pb}_{1-x}\mathrm{Sn}_x\mathrm{Te}$~\cite{NimtzSchlicht_1983_LeadSaltMonograph,Dimmock1966_PbSnTeInversion,Dimmock_1971Monograph}, and that reducing the gap makes the system approach an ideal Dirac electron system~\cite{AkibaTokunaga_PbTe2018,Izaki-PiMatrix}.
In this study, we specifically examine the cases of $x = 0.0$ ($\Delta = 90~\mathrm{meV}$) and $x = 0.35$ ($\Delta = 7~\mathrm{meV}$), in order to investigate how the degree of Dirac-like character influences the magnetic response.

We calculated the Landau levels of $\mathrm{Pb}_{1-x}\mathrm{Sn}_x\mathrm{Te}$ by combining the Lent tight-binding model and the $\pi$-matrix method for the $\mathrm{L}$ point, where the conduction and valence bands meet.
The dimension of the $\pi$-matrix was set to $N = 300$.
We assumed the magnetic field lies along the $c$-axis of the crystal.

\begin{figure}[tp]
    \centering
    \includegraphics[width=0.85\columnwidth]{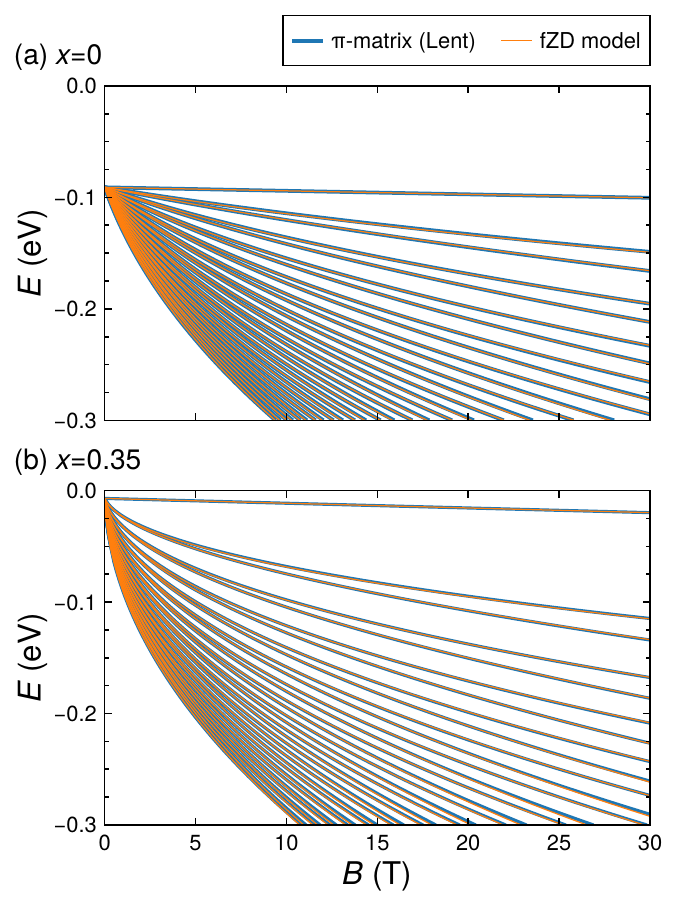}
    \caption{Magnetic-field dependence of Landau levels at the L point for (a) $x=0$ (PbTe) and (b) $x=0.35$ with $\kappa_{\mathrm{soc}}=1$.
Blue thick lines show the $\pi$-matrix results, and orange thin lines indicate the fitted fZD model.}
    \label{fig:landau_levels}
\end{figure}

Figures~\ref{fig:landau_levels}(a) and (b) show the magnetic-field dependence of the Landau levels at the $\mathrm{L}$ point for $\kappa_{\mathrm{soc}} = 1$, corresponding to $x = 0$ and $x = 0.35$, respectively.
The orange thin lines represent the results of the fZD model fitted to the $\pi$-matrix data (blue thick lines).
The excellent agreement between the two validates the applicability of the fZD model for analyzing the magneto-electronic properties of $\mathrm{Pb}_{1-x}\mathrm{Sn}_x\mathrm{Te}$.

Based on the fitted fZD model [equation~(\ref{eq:fZD_model})], we calculated the magnetization using equations~(\ref{eq:magnetization}) and (\ref{eq:thermodynamic_potential}).
Since $\mathrm{Pb}_{1-x}\mathrm{Sn}_x\mathrm{Te}$ possesses four valleys at the $\mathrm{L}$ points, we summed up the contributions from all the valleys.
To ensure convergence of the summation over Landau levels, we employed the cutoff function [equation~(\ref{eq:cutoff_function})].

\begin{figure}[tp]
    \centering
    \includegraphics[width=0.95\columnwidth]{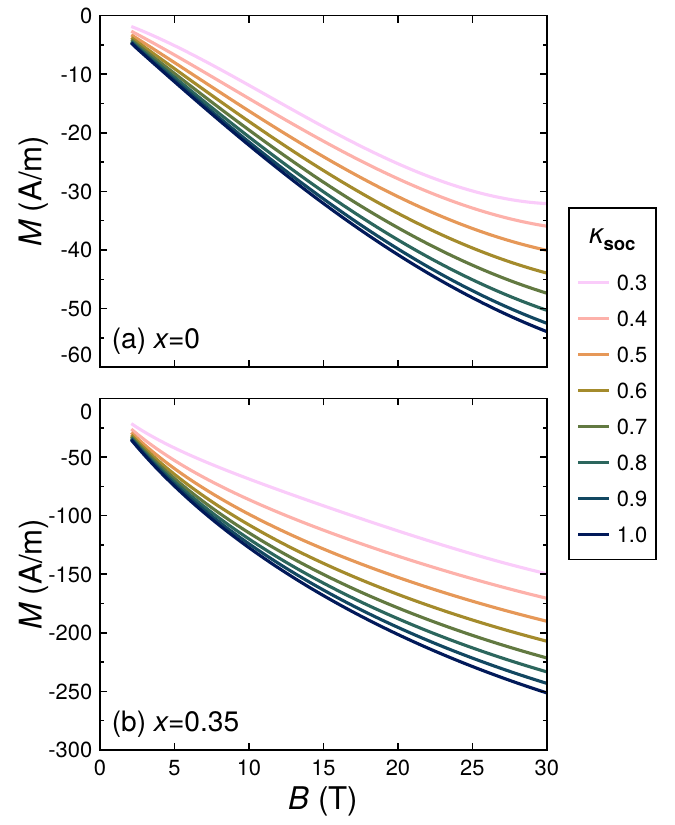}
    \caption{Magnetic field dependence of magnetization in $\mathrm{Pb}_{1-x}\mathrm{Sn}_x\mathrm{Te}$ with varying SOC strength for (a) $x=0$ and (b) $x=0.35$.}
    \label{fig:magnetization_pbsnte}
\end{figure}

Figure~\ref{fig:magnetization_pbsnte} shows the magnetic-field dependence of magnetization for (a) $x=0$ and (b) $x=0.35$ with varying the amplitude of SOC.
We emphasize that these results represent the first computation of macroscopic diamagnetic response based on material-specific electronic states obtained via the $\pi$-matrix method.
This approach provides a significant advantage over previous studies of Dirac electron systems~\cite{Fukuyama2007_2DWeylDiamagnetism,KoshinoAndo2010_DiamagnetismGrapheneWolff,FuseyaOgataFukuyama2014_SHEDiamagnetism_AnisotropicDirac,KawamuraFuseya2023}, which typically relied on simplified effective models rather than realistic band structures.
In figure~\ref{fig:magnetization_pbsnte}, the calculation parameters were as follows:
chemical potential $\mu = 0~\mathrm{eV}$, temperature $T = 77~\mathrm{K}$, cutoff energy $E_{\mathrm{c}} = 0.21~\mathrm{eV}$, and smoothing parameter $\beta_{\mathrm{c}} = 8~\mathrm{eV}^{-1}$.
The summation in equation~(\ref{eq:thermodynamic_potential}) was performed over Landau levels up to $n \leq 2000$.
The results show that the magnetization is negative ($M < 0$) for all values of $\kappa_{\mathrm{soc}}$, indicating diamagnetic behavior.
The magnitude $|M|$ increases monotonically with the magnetic field $B$.
In the weak-field limit, $M \propto B$, while in the high-field region, the rate of increase diminishes.
This trend is consistent with previous reports for PbTe~\cite{AkibaTokunaga_PbTe2018}, Bi~\cite{Iwasa2019}, and other Dirac electron systems~\cite{KawamuraFuseya2023, PratamaUkhtarySaito2021_MagnetizationMassiveDirac, KeserLyanda-GellerSushkov2022_QEDDirac}.
Moreover, the magnitude of $|M|$ increases with increasing $\kappa_{\mathrm{soc}}$, indicating a clear connection between SOC strength and orbital diamagnetism.
For a given $\kappa_{\mathrm{soc}}$, the magnetization is also larger in the $x = 0.35$ case (with a narrower gap) compared to the $x = 0$ case.
The ratio between the two is approximately 5, highlighting the enhanced magnetic response in more Dirac-like systems.

\begin{figure}[tp]
    \centering
    \includegraphics[width=0.95\columnwidth]{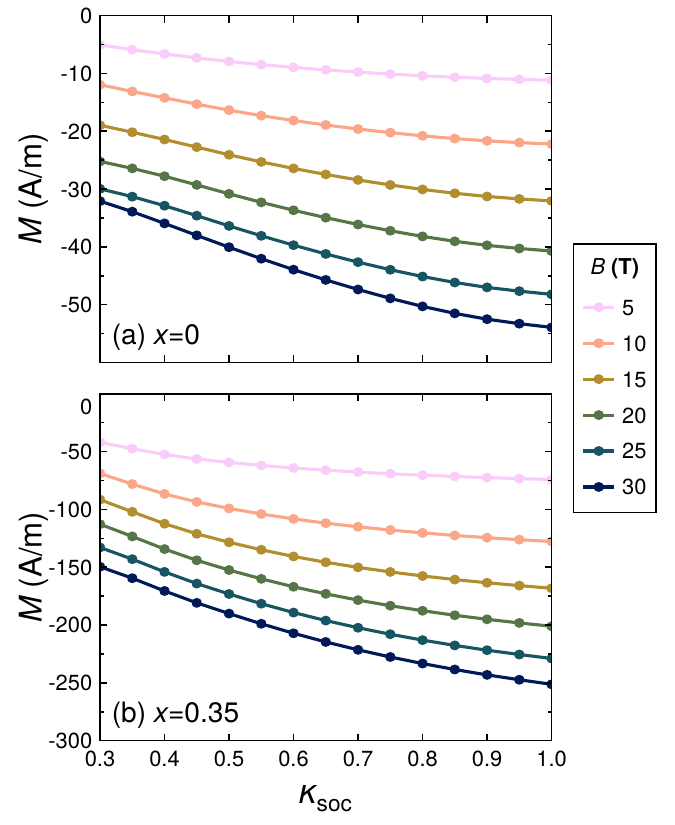}
    \caption{Magnetization as a function of SOC scaling parameter $\kappa_{\mathrm{soc}}$ in $\mathrm{Pb}_{1-x}\mathrm{Sn}_x\mathrm{Te}$ for (a) $x=0$ and (b) $x=0.35$.}
    \label{fig:vsSOC}
\end{figure}

The impact of SOC becomes even more evident when the magnetization is plotted as a function of $\kappa_{\mathrm{soc}}$.
As shown in figure~\ref{fig:vsSOC}, $|M|$ increases monotonically with increasing $\kappa_{\mathrm{soc}}$ for both $x = 0$ and $x = 0.35$.
This result clearly demonstrates that SOC plays a decisive role in enhancing orbital diamagnetism.
Indeed, this result provides a direct answer to the central question posed in the Introduction.

Furthermore, the slope $|M|/\kappa_{\mathrm{soc}}$ becomes steeper at higher magnetic fields, indicating that the contribution of SOC to magnetization is amplified as the field increases.
This behavior strongly suggests that SOC affects magnetization predominantly through interband magnetic effects.

In addition to SOC, we examine the influence of the anisotropy of the Fermi surface on the magnetization.
For a Dirac electron system with $\bm{B}\parallel z$, the magnetization $M$ is derived as follows: $M\propto -\left(\alpha_{xy}^2-\alpha_{xx}\alpha_{yy}\right)/\sqrt{\det\hat{\alpha}/\mathit{\Delta}}$ , where $\hat{\alpha}$ denotes the inverse mass tensor~\cite{FuseyaOgataFukuyama2014_SHEDiamagnetism_AnisotropicDirac}, which means that the magnitude of diamagnetism is proportional to the Gaussian curvature of the energy dispersion.
Within our current framework using $\pi$-matrix and LCAO models, it is hard to independently control the anisotropy of the Fermi surface.


\subsection{Analysis Based on the fZD Model}

The SOC dependence of magnetization can be systematically understood through the coefficients of the fZD model. As described in section~\ref{Sec_fZD}, each term in the fZD model contributes differently to the magnetic response: the free-electron term $C_{\mathrm{f}}$ and the Dirac-type interband coupling $C_{\mathrm{D}}$ produce diamagnetic contributions, while the Zeeman term $C_{\mathrm{Z}}$ contributes to paramagnetism. Thus, the net magnetization reflects the relative magnitudes of these competing effects.

\begin{figure}[tp]
    \centering
    \includegraphics[width=0.9\columnwidth]{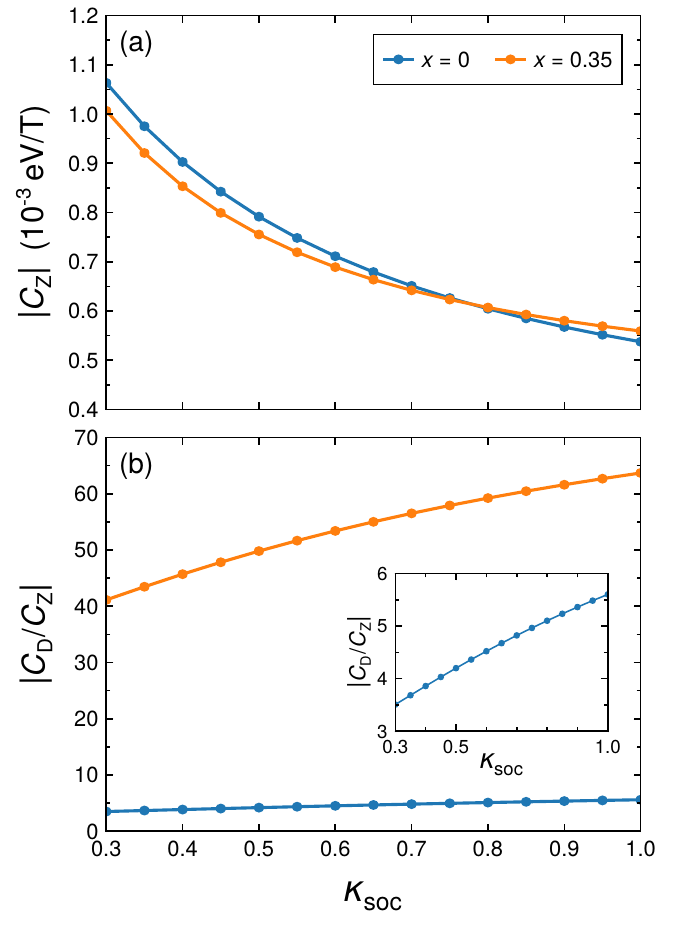}
    \caption{Dependence of fZD model parameters on the SOC scaling factor $\kappa_{\mathrm{soc}}$ for (a) Zeeman coefficient $C_{\mathrm{Z}}$ and (b) Ratio $|C_{\mathrm{D}}/C_{\mathrm{Z}}|$. The inset in (b) shows a magnified view for $x=0$.}
    \label{fig:fit_CZ_CDoverCZ_SOC}
\end{figure}

To clarify how SOC modifies these contributions in $\mathrm{Pb}_{1-x}\mathrm{Sn}_x\mathrm{Te}$, we extracted the fZD coefficients by fitting the model to the Landau levels computed using the $\pi$-matrix method.
Figure~\ref{fig:fit_CZ_CDoverCZ_SOC}(a) shows that the Zeeman term $C_{\mathrm{Z}}$ (note $C_{\mathrm{Z}}<0$) decreases monotonically in magnitude as $\kappa_{\mathrm{soc}}$ increases, with nearly identical behavior for $x = 0$ and $x = 0.35$.
In contrast, figure~\ref{fig:fit_CZ_CDoverCZ_SOC}(b) demonstrates that the ratio $|C_{\mathrm{D}}/C_{\mathrm{Z}}|$ increases steadily with $\kappa_{\mathrm{soc}}$, and that this ratio is consistently larger for $x = 0.35$ than for $x = 0$, indicating a more pronounced Dirac-type contribution in the narrower-gap regime.

These trends highlight a key result: although SOC is often expected to enhance both $C_{\mathrm{Z}}$ (via increased effective $g$-factors) and $C_{\mathrm{D}}$ (through stronger interband coupling), our analysis reveals a distinct imbalance—SOC suppresses the paramagnetic Zeeman term while enhancing the relative influence of the interband Dirac term.
As demonstrated in figure~\ref{fig:magnetization_heatmap}, such an increase in $|C_{\mathrm{D}}/C_{\mathrm{Z}}|$ leads to a stronger diamagnetic response.

This finding provides a microscopic explanation for the enhancement of orbital diamagnetism with SOC observed in our material-specific calculations.
It clarifies that the dominant mechanism behind this enhancement is not an increase in spin splitting, but rather a SOC-induced amplification of the Dirac-type interband effect.

Significantly, this result bridges relativistic electronic structure and macroscopic magnetization, offering a clear and physically grounded understanding of SOC-enhanced diamagnetism.
The fZD model thus serves not only as a practical fitting tool but also as a conceptual framework for disentangling the competing contributions to magnetism in multiband, narrow-gap semiconductors.

We now comment on the origin of the Zeeman term emerging in our analysis.
Those Zeeman terms all originate from the SOC.
Since the $g$-factor for $\mathrm{Pb}_{1-x}\mathrm{Sn}_x\mathrm{Te}$ is known to be 60 to several hundreds~\cite{HayasakaFuseya2016_SOCZeemanSplittingPbSnTe} while that for the bare electron is 2, the contribution to the Zeeman terms from the spin magnetic moment of the bare electron is negligible in the material.
In the current calculation, the effect of SOC is directly incorporated through $\lambda_{\rm c, a}$ in the relativistic LCAO Hamiltonian by Lent \textit{et al.}~\cite{Lent_PbTeBands}.

\subsection{Connection to experiment}

\begin{figure}[tp]
    \centering
    \includegraphics[width=0.95\columnwidth]{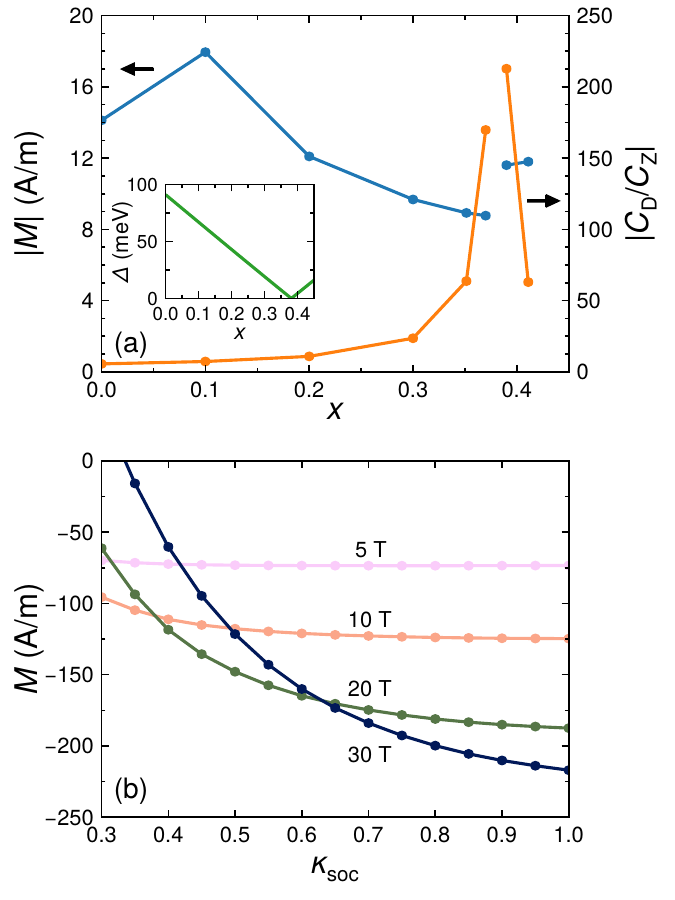}
    \caption{
    (a) $x$ dependence of diamagnetism at $B=5\,\mathrm{T}$ with $\mathit{\mu}=-70\,\mathrm{meV}$ and $\kappa_{\rm soc}=1$ (left axis), $|C_{\mathrm{D}}/C_{\mathrm{Z}}|$ (right axis) and energy gap parameter $\mathit{\Delta}$ (inset).
    (b) $\kappa_{\mathrm{soc}}$ dependences of magnetization for $x=0.41$.}
    \label{fig:variation_on_x}
\end{figure}

We have established that SOC enhances orbital diamagnetism.
To enable experimental verification of this conclusion, we examine the $x$ (composition) dependence in $\mathrm{Pb}_{1-x}\mathrm{Sn}_x\mathrm{Te}$, which is experimentally controllable to adjust SOC in real alloy systems~\cite{ChiuSinghMardanyaBansil2020_TopoTransitionNa3Bi,PulkkinenKothalawalaSuzukiBansil2025_NaSbBiTransisitionCompton}.
As $x$ increases, both the band gap $2\mathit{\Delta}$ and the SOC strength decrease in $\mathrm{Pb}_{1-x}\mathrm{Sn}_x\mathrm{Te}$~\cite{Lent_PbTeBands,Dimmock_1971Monograph}.
Those two effects compete with each other because a smaller band gap enhances diamagnetism, whereas a weaker SOC reduces it.

Figure~\ref{fig:variation_on_x}(a) shows the $x$-dependence of the magnetization at $B=5\,\mathrm{T}$ with $\kappa_{\mathrm{soc}}=1$ and realistic chemical potential $\mathit{\mu}=-70\,\mathrm{meV}$ on the left axis.
We first focus on the region with $x<x_c$, where the band gap closes at $x_c=0.38$.
In this region, $|M|$ decreases with increasing $x$ (except for $x\le 0.1$).
The trend in figure~\ref{fig:variation_on_x}(a) indicates that the SOC-originated suppression of $|M|$ outweighs the band-gap-driven enhancement, resulting in an overall reduction of diamagnetism as $x$ increases. This provides a concrete experimental test of our central claim, SOC-enhanced diamagnetism, by investigating $x$-dependence of $M$.
Note that the behavior for $x\le 0.1$ is different from that in the other region because the chemical potential $\mu$ lies within the band gap for the range of $x$.

As shown on the right axis of figure~\ref{fig:variation_on_x}(a), $|C_{\mathrm{D}}/C_{\mathrm{Z}}|$ increases as the band gap decreases.
This result indicates that the Dirac approximation becomes more accurate for smaller band gaps and that $|C_{\mathrm{D}}/C_{\mathrm{Z}}|$ strongly depends on the band gap.
In contrast to the $|M|$, $|C_{\mathrm{D}}/C_{\mathrm{Z}}|$ monotonically increases with $x$, since the value does not depend on the chemical potential $\mu$.

For $x>x_c$, the system lies in the topologically non-trivial phase~\cite{Xu2012_TCIPbSnTeARPES}.
Figure~\ref{fig:variation_on_x}(b) shows the $\kappa_{\mathrm{soc}}$ dependence of the magnetization for $x=0.41$, where the band gap is the same as that for $x=0.35$ (trivial phase).
The enhancement of the diamagnetic response with increasing $\kappa_{\mathrm{soc}}$ is preserved in the non-trivial region.
As shown in figure~\ref{fig:variation_on_x}(a), $|M|$ increases with $x$ in this region. 
In addition, the magnetic field dependence of the magnetization for $x=0.41$ [figure~\ref{fig:variation_on_x}(b)] differs from that for $x<x_c$ (figure~\ref{fig:magnetization_pbsnte}).
Topological transition might be one of the causes for the contrast in the behavior of the magnetizations.
A detailed discussion from a topological point of view is beyond the scope of this paper, and we leave it for future work.

At the critical point $x=x_c$, where the band gap reaches zero, diamagnetism is expected to diverge when $T=0$~\cite{FuseyaOgataFukuyama2014_SHEDiamagnetism_AnisotropicDirac}.
However, we can smoothly calculate the magnetization even at $x=x_c$, since the divergent behavior can be suppressed with $T>0$.

\begin{figure}[tp]
    \centering
    \includegraphics[width=0.95\columnwidth]{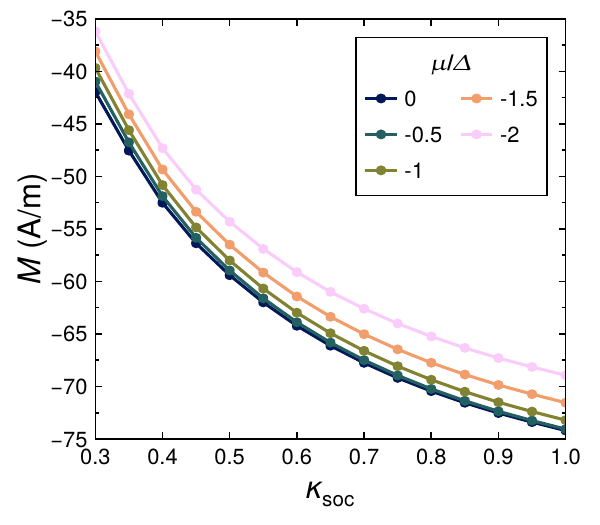}
    \caption{$\kappa_{\mathrm{soc}}$ dependences of magnetization with various chemical potentials $\mu$ for $x=0.35$.}
    \label{fig:variation_on_mu}
\end{figure}

We next discuss the effect of carrier doping by changing the chemical potential $\mu$.
Figure~\ref{fig:variation_on_mu} shows the $\kappa_{\mathrm{soc}}$ dependence of the magnetization for $x=0.35$ with several values of chemical potentials $\mu$.
The enhancement of the diamagnetic response with increasing $\kappa_{\mathrm{soc}}$ persists for all the $\mu$ considered.
The result also shows that the diamagnetism decreases monotonically with increasing $|\mu|$.
Since the density of carriers is of the order of $10^{18}\,\mathrm{cm^{-3}}$ in $\mathrm{Pb}_{1-x}\mathrm{Sn}_x\mathrm{Te}$~\cite{Allgaier1958_MRPbX}, which corresponds to the chemical potential on the order of $10\,\mathrm{meV}$, $|\mu| \simeq \Delta \sim 2 \Delta$ is in the realistic range ($\Delta =7$ meV for $x=0.35$).

An investigation into the temperature dependence of diamagnetism will be an interesting subject and substantially complement the present work.
Intuitively, while the magnitude of diamagnetism is expected to decrease with increasing temperature, the enhancement due to SOC holds even at elevated temperatures.
A reliable calculation of the temperature dependence is challenging due to technical limitations at this stage and beyond the scope of the present work.
We therefore leave this issue for future study.

\section{Conclusion}

In this work, we have investigated the role of spin--orbit coupling (SOC) in orbital magnetism, focusing on the narrow-gap semiconductor $\mathrm{Pb}_{1-x}\mathrm{Sn}_x\mathrm{Te}$.
To quantitatively evaluate the material-specific magnetization, we employed the $\pi$-matrix method, which enables gauge-invariant and reliable calculations of Landau levels under magnetic fields, faithfully reflecting the electronic structure of the material.

We have quantitatively computed the orbital magnetization for $\mathrm{Pb}_{1-x}\mathrm{Sn}_x\mathrm{Te}$, taking complete account of the intrinsic band structure and interband magnetic effects.
The results show that both compositions, $x = 0$ and $x = 0.35$, exhibit diamagnetic responses, with the magnitude of diamagnetism being larger for $x = 0.35$ due to its smaller band gap.
Furthermore, we have found that the diamagnetic response increases systematically with the SOC scaling factor $\kappa_{\mathrm{soc}}$.
This dependence clearly indicates that spin--orbit coupling enhances orbital diamagnetism.
Notably, the enhancement becomes more pronounced under stronger magnetic fields, suggesting that SOC contributes to diamagnetism primarily through interband magnetic field effects.

To clarify the underlying mechanism by which SOC influences orbital magnetism, we have introduced the free--Zeeman--Dirac (fZD) model and demonstrated that it successfully reproduces the Landau level structure near the band edge.
Analysis of the fitted fZD coefficients has revealed that increasing SOC leads to a monotonic suppression of the Zeeman term $C_{\mathrm{Z}}$ and an enhancement of the ratio $|C_{\mathrm{D}}/C_{\mathrm{Z}}|$.
This imbalance explains the observed enhancement of diamagnetism and highlights the importance of interband effects---rather than spin splitting---in governing the magnetic response under strong SOC.

Overall, this study provides a framework for understanding orbital magnetism in narrow-gap semiconductors by combining accurate band structure calculations with an analytically tractable effective model.
Our approach is not limited to $\mathrm{Pb}_{1-x}\mathrm{Sn}_x\mathrm{Te}$ but can be applied to a wide range of multiband systems where relativistic effects are significant.

These insights directly address the longstanding question posed in the introduction:
``Is spin--orbit coupling essential for orbital diamagnetism?"
Our answer, grounded in both numerical and analytical frameworks, is a resounding yes---``SOC fundamentally enhances orbital diamagnetism via the Dirac-type coupling term."


\ack
We thank M. Tokunaga and K. Akiba for valuable discussions.
This work was supported by JSPS KAKENHI
(Grants No. 23H04862, No. 23H00268, and No. 22K18318).

\section*{References}
\bibliographystyle{naturemag}
\bibliography{pbtemagne}

\end{document}